\documentclass[ijoc,nonblindrev]{informs3}
\OneAndAHalfSpacedXII %

\usepackage{latexsym,upref} %
\usepackage{xcolor}

\usepackage{natbib}
 \bibpunct[, ]{(}{)}{,}{a}{}{,}%
 \def\bibfont{\small}%

\usepackage{float}
\usepackage{enumerate}
\usepackage{dsfont}
\usepackage{xfrac}

\usepackage{tikz}
\usetikzlibrary{arrows.meta}
\usetikzlibrary{backgrounds}

\usepackage{pgfplots}
\pgfplotsset{compat=newest}
\usepgfplotslibrary{groupplots}
\usepgfplotslibrary{polar}
\usepgfplotslibrary{smithchart}
\usepgfplotslibrary{statistics}
\usepgfplotslibrary{dateplot}
\usepgfplotslibrary{ternary}
\usepgfplotslibrary{patchplots}
\usepgfplotslibrary{fillbetween}
\pgfplotsset{%
layers/standard/.define layer set={%
    background,axis background,axis grid,axis ticks,axis lines,axis tick labels,pre main,main,axis descriptions,axis foreground%
}{grid style= {/pgfplots/on layer=axis grid},%
    tick style= {/pgfplots/on layer=axis ticks},%
    axis line style= {/pgfplots/on layer=axis lines},%
    label style= {/pgfplots/on layer=axis descriptions},%
    legend style= {/pgfplots/on layer=axis descriptions},%
    title style= {/pgfplots/on layer=axis descriptions},%
    colorbar style= {/pgfplots/on layer=axis descriptions},%
    ticklabel style= {/pgfplots/on layer=axis tick labels},%
    axis background@ style={/pgfplots/on layer=axis background},%
    3d box foreground style={/pgfplots/on layer=axis foreground},%
    },
}

\newcommand\junk[1]{}
\newcommand\arXiv[1]{{\tt arXiv:\href{https://arxiv.org/abs/#1}{#1}}}
\newtheorem{definition}{Definition}
\newtheorem{problem}{Problem}
\junk{
\theoremstyle{plain}
\newtheorem{theorem}{Theorem}[section]

\newtheorem{problem}[theorem]{Problem}

\newcommand{\old}[1]{}
\theoremstyle{definition}
\newtheorem{definition}[theorem]{Definition}

    }

\def\P{\mathcal{P}}
\def\I{\mathcal{I}}
\def\J{\mathcal{J}}

\usepackage[ruled,section]{algorithm}
\usepackage{algpseudocode}
\algnewcommand{\IIf}[1]{\State\algorithmicif\ #1\ \algorithmicthen}
\algnewcommand{\EndIIf}{\unskip\ \algorithmicend\ \algorithmicif}
\algnewcommand{\FForAll}[1]{\State\algorithmicforall\ #1\ \algorithmicdo}
\algnewcommand{\EndFFor}{\unskip\ \algorithmicend\ \algorithmicfor}
\algblock[indent]{BeginIndent}{EndIndent}{}{}

\usepackage{array}
\usepackage[font=scriptsize]{subcaption}
\usepackage{textcomp}
\usepackage{stfloats}
\usepackage{verbatim}
\usepackage{balance}

\definecolor{color1}{HTML}{009AFA}
\definecolor{color2}{HTML}{E36F47}

\usepackage{hyperref}
\usepackage{cleveref}
\Crefname{problem}{Problem}{Problems}

\begin{document}
\RUNTITLE{Minimizing interference-to-signal ratios in multi-cell telecommunication networks}
\RUNAUTHOR{P.L. Erdős \& T.R. Mezei}
\TITLE{Minimizing interference-to-signal ratios in multi-cell networks}
\ARTICLEAUTHORS{%
\AUTHOR{Péter L.\ Erdős, Tamás Róbert Mezei}
\AFF{Alfréd Rényi Institute of Mathematics, Loránd Eötvös Research Network, Budapest, Hungary, \\ \EMAIL{erdos.peter@renyi.hu}, \URL{https://orcid.org/0000-0002-1139-2316} \\ \EMAIL{mezei.tamas.robert@renyi.hu}, \URL{https://orcid.org/0000-0002-7608-3215}}
    }

\ABSTRACT{%
    In contemporary wireless communication networks, base-stations are organized
    into coordinated clusters (called \emph{cell}s) to jointly serve the users.
    However, such fixed systems are plagued by the so-called cell-edge problem:
    near the boundaries, the interference between neighboring clusters can
    result in very poor interference-to-signal-power ratios. To achieve a high
    quality service, it is an important objective to minimize the  sum of these
    ratios over the cells.

    The most common approach to solve this minimization problem is arguably the
    spectral clustering method. In this paper, we propose a new clustering
    approach, which is deterministic and computationally much less demanding
    than current methods. Simulating on synthetic instances indicates that our
    methods typically provide higher quality solutions than earlier methods.

    An earlier version of this algorithm was reported in \arXiv{2111.00885}
}%

\maketitle
\section{Introduction}\label{sec:intro}
One of the most fundamental pillars of modern life is
telecommunication in general, and wireless telecommunication networks in
particular. These serve literally billions of requests every week, and not only
for phones, but also devices from the Internet of Things (IoT). In this paper we
will refer to all these different units as \emph{user}s.

The first wireless networks were constructed around 1970, and were envisaged to
be built from small \emph{cell}s, each cell served by one \emph{base-station} (\emph{BS}
for short), so the network was decomposed into smaller parts which served the
users independently from each other. (At this time the users were bulky
\emph{mobile} phones; see, for example~\cite{tse}.) While this
idea is simple, the design is afflicted by the well-known \emph{cell-edge
problem}: users located in the overlapping area of two cells would suffer from
strong interference from the neighboring BSs. This problem has created a
considerable challenge for service providers.

In the current \emph{coordinated multipoint (CoMP)} transmission technology,
several (typically physically close) base-stations are organized, permanently or
dynamically, into one \emph{cell}, and the BSs in the same cell jointly serve all the
users in this cell. The goal of this approach is again to decompose a large-scale
network into smaller parts: the constructed \emph{clusters} (called again
\emph{cells}) have smaller engineering complexity than the original network and
they can operate in parallel.

Nowadays, considering the enormous number  of base-stations and users, in spite
of the multi-cell approach,  the cell-edge problem stubbornly remained with us
(see~\cite{gesb}). For 5G networks the cell-edge problem
can become even more pronounced because more users fall into the
cell-edge area as the size of each cell shrinks (see~\cite{dai}).

Mathematically speaking, minimizing the cell-edge problem belongs to the
family of \emph{clustering problems}: given a set of (not-necessarily) homogeneous
objects, we want to divide the objects into pairwise disjoint classes while
optimizing some ``measure'' of the resulting set of classes.

Myriads of theoretical and practical problems belong to the clustering
framework. They come from classical combinatorial optimization problems to printed circuit board design, from VLSI CAD applications to distributing tasks among processors
for supercomputing processes, from pattern recognition to computer vision
problems and image retrieval. The computational complexity of these problems
vary from easy to very hard. For example, the minimum number of edges whose
deletion places two predefined vertices into separate components in a simple
graph (\emph{minimum cut problem} ~\cite{menger}) can be computed in polynomial time. The situation changes dramatically if we want to separate three predefined vertices (\emph{the multiway cut problem}). In the general case this generalization of the problem was shown to be NP-hard (\cite{Dahl}), while the problem becomes fixed parameter tractable if the input is restricted to planar graphs (\cite{Dahl} and \cite{gold}).

Probably the very first engineering problems of such clustering nature was the
following: we want to place complicated electronic design on printed circuit
boards, where each board can contain at most $k$ components and where the
electronic connections among the boards are expensive compared to connections
inside a board (\cite{kern}, see also \cite{dorn}). It can be seen that in such a real-world problem, an upper bound on the possible size of the clusters is given, too. While the notion of NP-completeness was being developed at the time, the authors correctly placed the problem into the NP-hard class.

The majority of the clustering problems are NP-hard, so there is no chance to
solve them exactly. Sometimes there are known performance guarantees on the
solution: for example in~\cite{Dahl} there is a
polynomial time algorithm with an $2(1-1/k)$ approximation ratio for the
multiway-cut problem, while in \cite{Cali} a polynomial time $(1.5-1/k)$-approximation algorithm was developed for the same problem.

The interference minimization problem in wireless networks is a known NP-hard
problem, and so far a constant factor approximation algorithm
has not been found. In the literature there are several approaches to solve
this problem, see~\cite{akoum,dahr,huang,kara,tse,zhang}. Because of
the intractability of the problem, all of these approaches are heuristic in
nature. In practice, these methods still do not satisfactorily solve the
cell-edge problem~\cite{gesb}.

Essentially, all of these method use one of the general clustering methods: the
\emph{kernel $k$-means} or the \emph{spectral clustering} method. The former
method was developed in~\cite{macq} and~\cite{lloyd}. However, as it was proved in~\cite{dhillon04}, the two approaches are essentially equivalent with each other (for a survey on these methods, see~\cite{luxb}). Consequently, we will compare our
algorithm to the spectral clustering method (as it is used in~\cite{dai}). This approach attacks this clustering problem as an undivided one, partitioning the base-stations and users simultaneously. However, the two sets of agents typically have different cardinalities, and their elements have very different functions and properties
in the network.

\medskip\noindent

In this paper we propose a simple and fast clustering algorithm to deal with the
cell-edge problem. (An earlier version of this algorithm was reported in \cite{erdos}) Our algorithm runs significantly faster than the spectral clustering method, and simulations on synthetic instances indicate that our proposed method typically provides higher quality solutions than the application of the Spectral Clustering method in~\cite{dai}. In contrast with the spectral clustering methods, our proposed heuristic method is deterministic.

We divide the interference minimization problem into three subproblems. In the
first phase, a new, so-called \emph{dot-product similarity measure} is
introduced on pairs of base-stations. This similarity measure is based on the
dynamically changing signal-strengths between the users and the base-stations.
The two subsequent phases are two clustering problems. The second phase
partitions the base-stations into clusters, and the third phase assigns the
users to base-station clusters. The solution to the whole problem is given as a
pairing between the base-station clusters and user clusters.

Our reason for this handling is the following observation: the roles of the
base-stations and the users are different, and so they require different
considerations. We will emphasize this asymmetry with our notation system as
well. The clusters of the entire system will appear as pairs of clusters: one on the
index set of base-stations, and one on the index set of the users; cluster
classes of the same subscript will serve together as a cluster class of the
entire system.

The \emph{novelty} of our method, which is responsible for the superior
performance, lies in the usage of the dynamical similarity function between the
base-stations. The second and third phases may use off-the-shelf clustering
algorithms, which may allow possible further fine-tuning of our method.

\section{The total interference minimization problem}\label{sec:IMP}
The formulation of the \emph{total interference minimization problem} that we
will use in this paper was proposed and studied by Dai and Bai in their influential paper~\cite{dai}. We will give our description
of their formulation in the following paragraphs.

There is a collection $B:=\{b_i : i \in I\}$ of distinct base-stations and there
is a collection $U:=\{u_j : j\in J \}$ of distinct users, where $b=|B|$ and
$u=|U|$, and the base-stations and the users are indexed by the sets of the
first $b$ and $u$ natural numbers. As we mentioned earlier, the users can be
mobile phones, but can also be devices of the IoT. Therefore their numbers altogether can be rather large compared to the number of the base-stations. (However, in future 6G networks, the ratio of these numbers may change considerably. We do not consider this case here.)

The model depicts the network with a bipartite graph: one class contains the BSs while the other class consists of the users. Let $G=(V;E):=(B \cup U; E)$  where $E$ consists of ordered pairs of form $(b_i,u_j)$. We will use the shorthand $(i,j)$ as well.

We define a weight function $w : B\times U \rightarrow \mathbb{R}^+_0$, where
each $w_{i,j}$ is a positive real number if and only if $(i,j)$ is an edge in
$G$, otherwise we set $w_{i,j}=0$. The weight of an edge of the bipartite graph
represents the signal strength between its endpoints (one BS and one user).

Let $M$ be a positive integer, and let $[M]$ be the set of the integers from $1$ to $M$. Let $\I=(I_1,\ldots, I_M)$ and $\J=(J_1,\ldots,J_M)$ be partitions of $I$ and $J$, respectively. Finally, let $\P$ be the set of partition pairs: $\P:= \{P_\ell=(I_\ell,J_\ell) : \ell \in [M]\}$.

Let us define the following quantities: for each integer $\ell \in [M]$, let
\begin{equation}\label{eq:not1}
w(P_\ell):= \sum_{i\in I_\ell, j\in J_\ell} w_{i,j} \quad \text{and}  \quad
\bar w(P_\ell):= \sum_{i\in I_\ell, j\in J\setminus J_\ell} w_{i,j}+ \sum_{i\in I\setminus I_\ell, j\in J_\ell} w_{i,j}.
\end{equation}
In graph theoretical terms, the first quantity is the \emph{weight} of the partition class, while the second one is the \emph{cut value} of the partition class.

\begin{definition}[IF-cluster system]\label{def:cluster}
    A partition $\P$ is an \emph{IF-cluster system} (or IF-cluster for short; IF abbreviates \emph{interference}), if
    \begin{enumerate}[{\rm (i)}]
        \item there is no partition class $P_\ell$ such that $I_\ell=\emptyset$,
            and
        \item for any user, there is base-station in its cluster to which its joined by an edge in $G$.
    \end{enumerate}
    The \emph{total interference} (see~\cite{dai}) of a given $\P$ IF-cluster system is defined as
    \begin{equation}\label{eq:inf}
        \mathtt{tinf}(\P) = {\sum_{\ell\in[M]}}^{\displaystyle *}\  \frac{\bar w(P_\ell)}{w(P_\ell)},
    \end{equation}
    where the star superscript denotes that if a partition class $J_k$ is empty, then the index $\ell$ skips $k$, so that the formula in~\eqref{eq:inf} is well-defined. \qed{}
\end{definition}
\noindent Condition (ii) above covers the requirement that all users must be
served in a solution. The omission of a partition class in the sum
(\ref{eq:inf}) is due to the technical ability that some base-stations that do
not serve any users can be switched off temporarily.

The main result of this paper is new and fast heuristic algorithm (the
\textsc{Dot-product clustering} algorithm) for the following problem.
\begin{problem}[Total interference minimization problem]\label{th:interference}
Find an IF-cluster system $\P$ which minimizes~\eqref{eq:inf}.
\end{problem}

\medskip\noindent
As we mentioned earlier, the majority of the clustering problems in general, and the total interference minimization problem in particular,  are NP-hard. However, our \Cref{th:interference} lives on bipartite graphs, so the complexity results on general clustering problems do not apply to it automatically. However, incidentally, \Cref{th:interference} is also NP-hard. There are a vast number of graph partitioning problems that are similarly naturally defined on bipartite graphs. For example, the typical machine learning and data mining applications, such as product recommendation in e-commerce,  topic modeling in natural language processing, etc., are all naturally represented on bipartite graphs.

So it is not surprising that already in 2001, in (\cite{dhillon01}) the spectral graph partitioning algorithm was  used to co-cluster documents and words
(in bipartite graphs). From that time on, the spectral clustering method is also
often applied to solve other bipartite partitioning problems as well.

\medskip\noindent As we mentioned earlier, the model above was introduced by
Dai and Bai (\cite{dai}). Their approach  was
static: they evaluate the input, then they cluster the base stations and the
users  simultaneously to minimize the total interference. For that end, the
spectral clustering method was applied to construct the cluster system. The
developed method solves a relaxed quadratic programming problem and constructs
the clusters by discretizing the continuous solution. If a derived solution
contains a partition class without base-stations, then the solution is
dismissed. This approach is static, since it does not provide an efficient
method to deal with small, dynamic changes as time passes.

However, this static approach leaves much to be desired, since we should
consider some additional objectives: for initialization of the base-station/user
clustering in our wireless communication network, we want a fast, centralized
algorithm, like our proposed algorithm for clustering for the total
interference minimization problem will be. Furthermore, during the routine operation of the network, dynamic changes may occur: some users may move away from the BSs of a
given cluster, some may finish calls, while others (currently not represented in
the bipartite graph) may initiate calls. While these changes can be managed in a
centralized fashion, this would not be practical. Instead we need an incremental
algorithm, that is able to adaptively change the edge weights and/or can update
the actual vertices, and can manage the re-clustering of the affected users. It
is propitious to manage these local changes distributively by the users.
Finally, every few seconds it is useful to run the centrally managed algorithm
again to find a new clustering solution. Since the proposed algorithm
has a low complexity, this approach is clearly beneficial. We will return to
this question at the end of \Cref{sec:runningtimes}.

Our proposed \textsc{DP-Similarity Clustering} algorithm can handle all these issues as well.
As the simulations in \Cref{sec:exp} show, it is fast and provides high quality solutions, compared to the spectral clustering algorithm.

\section{Dot-product clustering algorithm for total interference minimization problem}\label{sec:solve-erc}

In this section we describe our  new and simple heuristic algorithm for the total interference minimization problem. As we already mentioned, our algorithm runs in three phases. In the first a similarity function will be introduced. This phase contains the novelty of our approach. Our similarity measure depends on the relations between the base-stations and the users.

In the second phase the base-stations will be clustered on the basis of the similarity measure. Here we have significant freedom to choose our clustering algorithm. The simplest possible method is arguably a hierarchical clustering method. For simplicity we use such a method in this paper, but this choice may badly affect the stability of the solutions. It may provide unbalanced cluster sizes, and it can also introduce too much engineering complication. It is possible that some back-step or averaging approach can amend the variance of the quality of the solutions.

Finally the third phase will assign users to the base-station clusters. By design, the output of our algorithm will always be an IF-cluster system. However, it would be beneficial to study methods to balancing the number of users in the cluster classes.

\subsection{Phase 1: The dot-product similarity measure}\label{sec:dot}
A superficial study of \cref{eq:inf} says that we want to decompose the graph
in such a way that clusters contain \emph{heavy} (high weight) edges, and
the cuts among the clusters consist of \emph{light} edges. The weight
function is described via the matrix $W$ where the rows correspond to the BSs,
and the columns correspond to the users:
\begin{equation*}
    W={(w_{i,j})}_{i\in \I,j\in \J}
\end{equation*}
Let $w_{i,\bullet}$ denote the row of $b_i$ (the $i$th BS), and let $w_{\bullet,j}$
denote the column of $u_j$ (the $j$th user). Hence we can write $W =
{[w_{i,\bullet}]}_{i\in \I}={[w_{\bullet,j}]}_{j\in \J}$. Our heuristic is that
the higher the correlation between the weight distribution of two BSs, the more
beneficial to include them in the same cluster. We define the
\emph{similarity function} as
\begin{equation}\label{eq:dot}
	\rho : \I \times \I \rightarrow \mathbb{R}^{\ge 0} \quad \text{with} \quad
	\rho (i,k) := \frac{w_{i,\bullet}^T \cdot
	w_{k,\bullet}}{\|w_{i,\bullet}\|\cdot \|w_{k,\bullet}\|}
\end{equation}
among the BSs, where $\|\cdot\|$ is the Euclidean-norm. The name
\emph{dot-product} is in reference to the enumerator of \cref{eq:dot}. The
similarity $\rho$ depends only on the relations between the users and the BSs. We
envisage that the larger the value of $\rho$, the greater the similarity between
the BSs.

\medskip

In the total interference minimization model, an ensemble of BSs in a cluster behave as one base-station.
Indeed, if the IF-cluster system $\{P_1,\ldots,P_M\}$ minimizes \cref{eq:inf}, then
replacing the ensemble of BSs in cluster $I_\ell$ with just one new BS $b_\text{new}$ whose weight to user $j$ is $\sum_{k\in I_\ell}w_{k,j}$ preserves the optimum, and the total interference metric takes this optimum on the partition pair $\I',\J$ where the $\ell$th class $I_\ell$ is replaced with the index of $b_{\text{new}}$ in $\I '$. Let us define
\begin{equation}\label{eq:vec}
    \mathrm{vec}(I_\ell)=\sum_{i\in I_\ell}w_{i,\bullet}
\end{equation}
as the sum of the signal strength vectors of the base-stations in $B_k$. The
similarity function $\rho$ can be naturally extended to ensembles of BSs:
\begin{equation}\label{eq:dot2}
    \begin{split}
        &\rho : 2^I \times 2^I \rightarrow \mathbb{R}^{\ge 0} \quad \text{with}\\
        &\rho (I_k,I_m) := \frac{{\mathrm{vec}(I_k)}^T \cdot
            \mathrm{vec}(I_m)}{\|\mathrm{vec}(I_k)\|\cdot \|\mathrm{vec}(I_m)\|}.
    \end{split}
\end{equation}

\subsection{Phase 2: Defining BS clusters}\label{sec:BS}
As we discussed it earlier, we have great freedom to determine the BS clusters.
However, for simplicity, here we apply a \emph{hierarchical} clustering
algorithm: we call it \textsc{DPH-clustering}, short for \emph{dot-product
hierarchical clustering}. The fixed integer $M$ which is the size of the cluster
system is part of the input. At the beginning, to each BS we assign a cluster
containing it. Then we will recursively merge the two clusters that have the
highest similarity $\rho$ between them, until the prescribed number of clusters
is reached. As we will soon see, this works reasonably well. Here we want to
draw attention to the fact that using normalization in \cref{eq:dot,eq:dot2} is
a natural idea.

\medskip

We start with the finest partition: let $\I_0$ consist of the individual clusters for each BS in $I$ (thus $|I|=|\I_0|$). We combine two clusters in each of the $|I|-M$ rounds iteratively to derive a sequence of partitions $\I_0,\I_1, \ldots, \I_{|I|-M}$ of $I$,
where $|\I_r|=|I|-r$.  $\I_r$ is obtained from $\I_{r-1}$ by combining the two clusters of $\I_{r-1}$ with the largest similarity $\rho$ between them as defined by \cref{eq:dot2}.

In \Cref{alg:clusteringB} we will maintain the similarity measure $\rho(I_k,I_m)$ for every pair of clusters $I_k,I_m\in \I_r$ as follows. Let us define the symmetric function $\mathit{dot}$ for every $k=1,\ldots,M$ as
\begin{align}\label{eq:dotsum}
\mathit{dot}(I_k,I_m)={\mathrm{vec}(I_k)}^T \cdot \mathrm{vec}(I_m).
\end{align}
If $\mathit{dot}$ is already computed for every pair in $\I_r\times
\I_r$, then the similarity measure $\rho$ can be computed via three
scalar operations for any pair of clusters in $\I_r\times
\I_r$, since
\begin{align}\label{eq:rhodot}
    \rho(I_k,I_m)=\frac{\mathit{dot}(I_k,I_m)}{\sqrt{\mathit{dot}(I_k,I_k)\cdot \mathit{dot}(I_m,I_m)}}.
\end{align}

\begin{algorithm}
	\caption{Hierarchical clustering based on the similarity function $\rho$}\label{alg:clusteringB}
	\begin{algorithmic}
		\Function{DPH-clustering}{$I,W,M$}
        \State $\I_0\gets \binom{I}{1}$
        \State $\mathit{dot}\gets W\cdot W^T$ \Comment{matrix multiplication}
		\For{$r=0$ to $|I|-M-1$}
        \State $\displaystyle\{I',I''\}\gets\argmax_{\{I_k,I_m\}\in
            \binom{\I_r}{2}}
            \rho(I_k,I_m)$\Comment{\eqref{eq:rhodot}}
        \State $\I_{r+1}=\I_r-\{I',I''\}+\{I'\cup I''\}$%
        \State update $\mathit{dot}$ \Comment{$\mathcal{O}(b)$ scalar operations}
		\EndFor
        \State\Return $\I_{|I|-M}$
		\EndFunction
	\end{algorithmic}
\end{algorithm}

It is easy to see that the running time of \Cref{alg:clusteringB} is in the
range of $\mathcal{O}\left(b^2u+(b-M)\cdot b^2\right)$, because when two
clusters are combined, $\mathit{dot}$ can be updated by summing the
corresponding two rows and two columns. Moreover, we may store the $\rho$-values
of pairs in $\I_r\times \I_r$ in a \emph{max-heap}: when two
clusters are merged, at most $2b$ values need to be removed and at most $b$ new
values need to be inserted into the heap which contains the at most
$\binom{b}{2}$ elements of the set $\{\rho(I_k,I_m)\ |\ \{I_k,I_m\}\in
\binom{\I_r}{2}\}$. With these optimizations, the \emph{for}-loop
takes at most $\mathcal{O}\left((b-M)\cdot b\log b\right)$ steps, thus the
running-time of the algorithm is dominated by the matrix multiplication $w\cdot
w^T$. There are many techniques to accelerate the multiplication of  matrices,
which we do not discuss here.

\subsection{Phase 3: Assigning users to BS clusters}\label{sec:users}
Let $\I:=\I_{b-M}=\{I_1,\ldots,I_M\}$ be the final partition produced by
the hierarchical clustering on the primary class. We are looking for the desired
final graph partition in the form of $\P=\{ (I_k,J_k)\ |\ k\in [M]\}$, so it only remains to find an appropriate clustering  $\J=\{J_1,\ldots,J_M\}$ of $J$.

We assign each user $j\in J$ to the cluster $J_\ell$ where $\ell$ is defined by
\begin{equation}\label{eq:user}
	\ell=\argmax_{k\in [M]} \sum_{i\in I_k}w_{i,j}.
\end{equation}
The choice described by \cref{eq:user} is easy to compute. A high-level pseudo-code can be found in \Cref{alg:dpBbestU}. If no element of $U$ is isolated in $G$, then \Cref{alg:dpBbestU} ensures that there are no isolated users in the subgraph induced by the base-station of $I_k$ and users $J_k$. However, this is not necessarily the case for elements of $I_k$. But the definition of ${\sum}^*$ in \cref{eq:inf} takes care of this. This completes the description our algorithm. \qed{}
\begin{algorithm}
    \caption{Dot-product hierarchical clustering on $I$ then assigning each
    element of $J$ to the best cluster.}\label{alg:dpBbestU}
	\begin{algorithmic}
        \Function{DP-Similarity Clustering}{$G,W,M$}
        \State $I,J\gets$ index sets of $B$ and $U$
        \State $\{I_1,\ldots,I_M\}\gets \Call{DPH-clustering}{I,W,M}$
        \State $J_1,\ldots,J_M\gets $ empty clusters
        \ForAll{$j\in J$}
	        \State $\ell\gets\argmax_{k\in [M]} \sum_{i\in I_k}w_{i,j}$
            \State add $j$ to $J_\ell$
        \EndFor
        \State\Return$\{ (I_k, J_k)\ |\ J_k\neq\emptyset\}$
		\EndFunction
	\end{algorithmic}
\end{algorithm}

\subsection{Engineering complexity}\label{sec:complex}
In this subsection, we discuss in short some practical considerations in
real-life applications. One of the most  important ones is that the clusters cannot
be arbitrarily complex (from an engineering point of view), because the
computational overhead of the synchronization of too many base-stations.
Therefore it may be necessary to consider an upper bound $T$ on the possible
numbers of the BSs in any cluster.

The number of clusters is part of the input of the our clustering method. One
can ask whether there is a way to optimize $M$. A heuristic attempt is given
in~\cite{dai}, using a binary search wrapper over the spectral
clustering method to determine an optimal $M$. The idea is based on a theorem of
Dai and Bai that the optimal solution is monotone increasing
in $M$, however, no such guarantee is given for the approximate solution found
by the heuristic algorithms constructing the $M$-part clusters. For this reason,
we do not consider this alternate optimization problem.

\section{Experiments}\label{sec:exp}

We have compared the performance of \textsc{DP-Similarity~Clustering} (\Cref{alg:dpBbestU}), and \textsc{Spectral~Clustering}~\cite{dai} in several scenarios.
In each case, base-stations (BSs) and users are randomly uniformly distributed, and independently placed into ${[0,1000]}^2$ (a square with an area of $1\,\mathrm{km}^2$). The weight (or signal strength) between a BS and a user $b_i,u_j\in{[0,1000]}^2$ is set to
\begin{equation*}
    w_{i,j}=\left\{
        \begin{array}{ll}
            {\|\mathrm{dist}_{\min}\|}^{-\alpha} & \text{ if }
            \|b_i-u_j\|\le \mathrm{dist}_{\min},\\
            {\|b_i-u_j\|}^{-\alpha} & \text{ if }
            \mathrm{dist}_{\min}\le\|b_i-u_j\|\le \mathrm{dist}_{\max},\\
            0 & \text{ if }\mathrm{dist}_{\max}<\|b_i-u_j\|,
        \end{array}
        \right.
\end{equation*}
where $\mathrm{dist}_{\min}=1$, $\mathrm{dist}_{\max}=200$, and we set the
\emph{path attenuation (path loss) exponent} $\alpha=2$
(see~\cite[Section~2]{tse}). We assume that in real-world applications the
signal strength values are readily available.

\textsc{DPH-clustering} (\Cref{alg:clusteringB}) is guaranteed to return an $M$-part clustering of the
base-stations (if $M\le b$). Consequently, the output of \textsc{DP-Similarity Clustering}
(\Cref{alg:dpBbestU}) is always an IF-cluster system (\Cref{def:cluster}), because it assigns each
user to the cluster with the largest weight to the user. However, this is not
the case for \textsc{Spectral Clustering}, which, in certain scenarios, is very
likely to create a cluster with some users but zero base-stations;
see~\Cref{fig:plots50failed,fig:plots200failed}.

\subsection{Observations.}
\begin{figure}
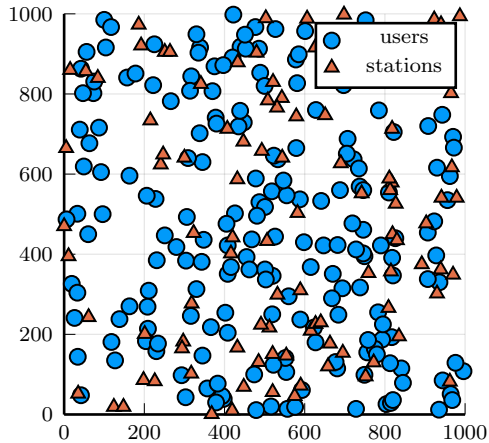
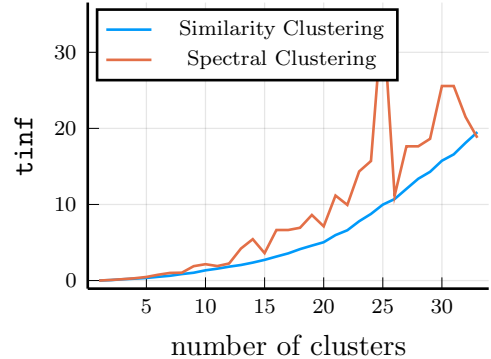
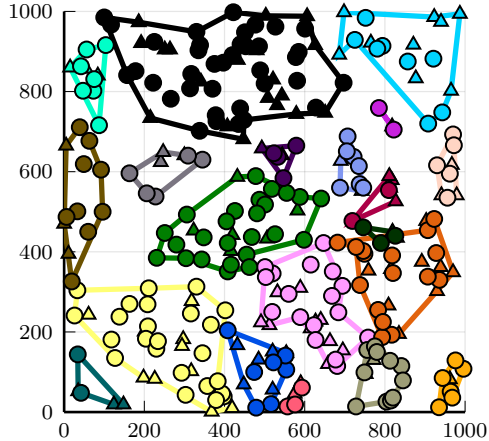
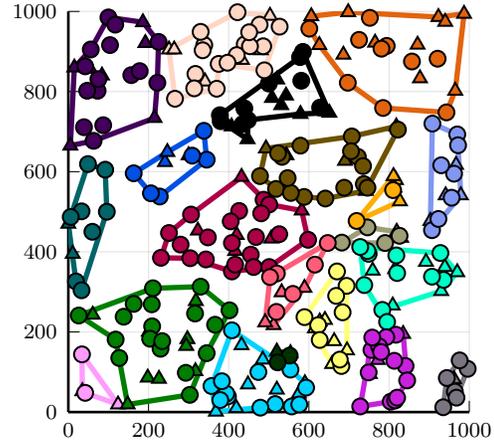
%
    \centering
    \begin{subfigure}[t]{0.47\textwidth}
        \centering

        \caption{100 BSs and 200 users, randomly and uniformly distributed}%
        \label{fig:comparison1:problem}
    \end{subfigure}\hfill
    \begin{subfigure}[t]{0.47\textwidth}
        \centering
%
        \caption{Comparison of the $\mathtt{tinf}$ values of the solution provided by the two algorithms as $M$ increases; \textsc{Spectral Clustering} failed for
        $M=27,30$ because it assigned a couple of users to a cluster without a
        base-station}%
        \label{fig:comparison1:scalingM}
    \end{subfigure}
    \begin{subfigure}{0.47\textwidth}
        \centering
%
        \caption{DP-Similarity Clustering into $M=20$ clusters, \Cref{alg:dpBbestU};
        $\mathtt{tinf}\approx 5.04$}%
        \label{fig:comparison1:similarity}
    \end{subfigure}\hfill
    \begin{subfigure}{0.47\textwidth}
        \centering
%
        \caption{Spectral Clustering into $M=20$ clusters, see~\cite{dai};
        $\mathtt{tinf}\approx 7.12$}%
        \label{fig:comparison1:spectral}
    \end{subfigure}
    \caption{A demonstration of the output of the two clustering algorithms on 100 BSs
    and 200 users. Clustering solutions shown for $M=20$ (so that the clusters
    are fairly large and visible). Triangles and circles
    represent BSs and users, respectively}\label{fig:comparison1}
\end{figure}
Let us start with comparing the two algorithms
(\textsc{DP-Similarity Clustering} and \textsc{Spectral Clustering}) on an
arbitrarily chosen clustering problem; after that, we will turn to a quantitative comparison.
\Cref{fig:comparison1:problem} shows a randomly and uniformly generated
placement of 100 BSs and 200 users. The scaling of the $\mathtt{tinf}$
(see~\cref{eq:inf}) of the solutions
provided by \textsc{Spectral Clustering} and \textsc{Spectral Clustering}
algorithm as a function of the number of clusters $M$ are shown on
\Cref{fig:comparison1:scalingM}.

\begin{figure}
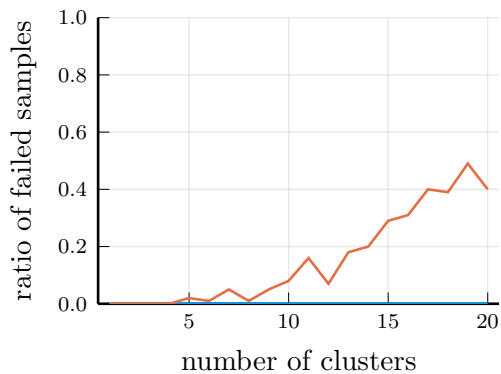
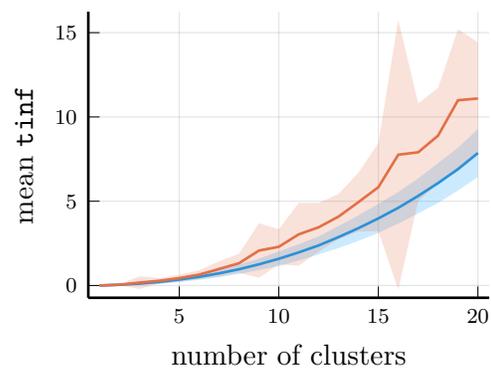
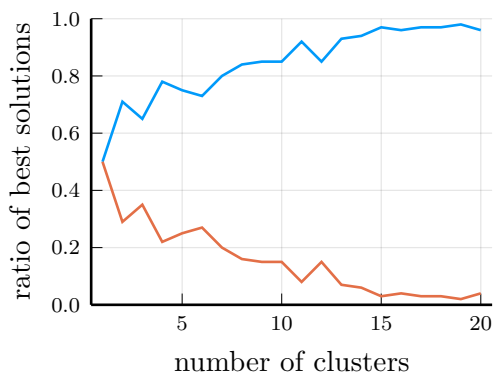
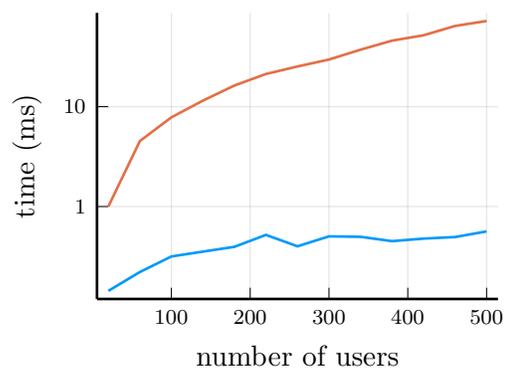

    \centering
    \begin{subfigure}[t]{0.47\textwidth}
        \centering

        \caption{For larger $M$, many of the solutions provided by
        \textsc{Spectral Clustering} have a cluster that contains some users,
        but zero BSs.}\label{fig:plots50failed}
    \end{subfigure}\hfill
    \begin{subfigure}[t]{0.47\textwidth}
        \centering
%
    \caption{Mean $\mathtt{tinf}$ (see~\cref{eq:inf}) values as a function of the number of clusters $M$; the ribbons show the standard deviation over the samples. Note, that the failed samples were not included in the sample mean and sample deviation.}
    \end{subfigure}
    \begin{subfigure}[t]{0.47\textwidth}
        \centering
%
        \caption{For each $M=1,\ldots,20$ the figure shows the ratios of those
            samples where \textsc{DP-Similarity Clustering} and where
            \textsc{Spectral Clustering} provided
            the solution with the smaller $\mathtt{tinf}$ value. Because of the
            high number of failures produced by \textsc{Spectral Clustering} for
            moderately large values of $M$, this ratio tends to $1$ for
    \textsc{DP-Similarity Clustering} (as $M$ increases).}
    \end{subfigure}\hfill
    \begin{subfigure}[t]{0.47\textwidth}
        \centering
%
        \caption{Running times for $b=50$ BSs and $M=20$ clusters, as the number of
        users grows; the $y$-axis is logarithmic.}
    \end{subfigure}
    \caption{Analyzing the output of
        \textcolor{color1}{\textsc{DP-Similarity Clustering}} and
        \textcolor{color2}{\textsc{Spectral Clustering}} for $b=50$ BSs and
        $u=200$ users;
        samples obtained over $100$ randomly generated
    problems.}\label{fig:plots50}
\end{figure}

\begin{figure}
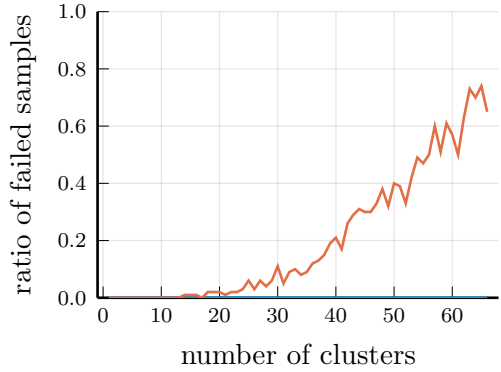
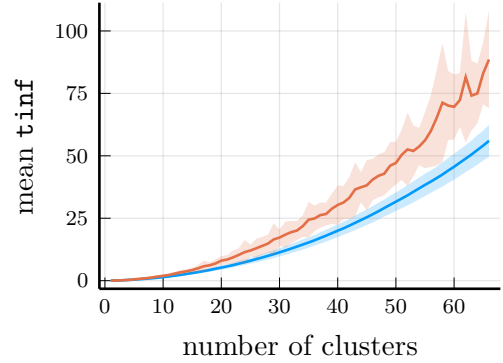
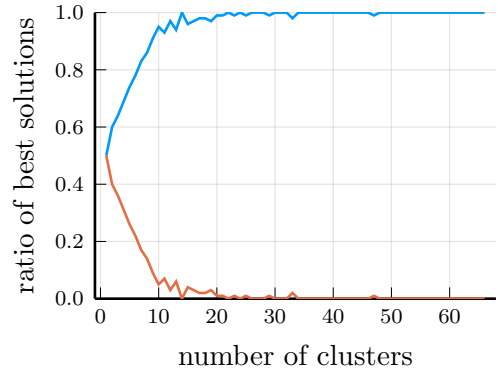
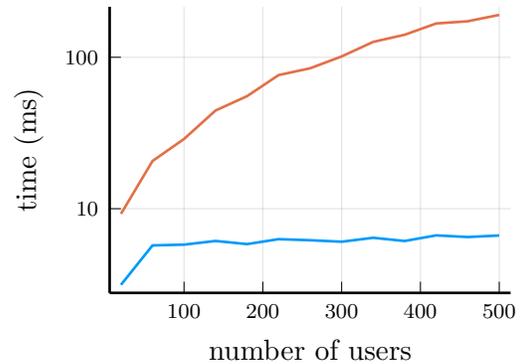

    \centering
    \begin{subfigure}[t]{0.47\textwidth}
        \centering
%
        \caption{For $M\ge 65$, more than half of the solutions provided by
        \textsc{Spectral Clustering} have a cluster that only contains users,
        but zero BSs.}\label{fig:plots200failed}
    \end{subfigure}\hfill
    \begin{subfigure}[t]{0.47\textwidth}
        \centering
%
        \caption{Mean $\mathtt{tinf}$ values as a function of the number of clusters
    $M$; the ribbons show the standard deviation over the successfully solved samples.}
    \end{subfigure}
    \begin{subfigure}[t]{0.47\textwidth}
        \centering
%
        \caption{For each $M=1,\ldots,66$ the graph shows the ratios of those
            samples where \textsc{DP-Similarity Clustering} and where
            \textsc{Spectral Clustering} provided
            the solution with the smaller $\mathtt{tinf}$ value. Although
            \textsc{DP-Similarity Clustering} provides the better solution in the
            majority of cases, it should be noted that the mean $\mathtt{tinf}$ values
            are about $30\%-50\%$ larger for the \textsc{Spectral
            Clustering} algorithm.}
    \end{subfigure}\hfill
    \begin{subfigure}[t]{0.47\textwidth}
        \centering
%
        \caption{Running times for $b=200$ BSs and $M=40$ clusters, as the number of
        users grows; the $y$-axis is logarithmic. Note that even for a moderate
        $u\approx 100$ users the running time of \textsc{Spectral Clustering}
        grows to $50\,\mathrm{ms}$, which may be prohibitive in
    applications.}\label{fig:plots200time}
    \end{subfigure}
    \caption{Analyzing the output of \textcolor{color1}{\textsc{DP-Similarity
        Clustering}} and \textcolor{color2}{\textsc{Spectral Clustering}}
        for $b=200$ BSs and $u=500$ users; samples obtained over $100$ randomly generated
        problems. Although \textsc{Spectral Clustering} performs slightly better
        then \textsc{DP-Similarity Clustering} for $M\le 26$, its running time is
        prohibitive in this regime of the total numbers of BSs and
        users.}\label{fig:plots200}
\end{figure}

\Cref{fig:plots50,fig:plots200} compare the performance of the algorithms for
$b=50$ and $200$ BSs. The plots correspond to the mean $\mathtt{tinf}$ values of
the solutions provided by the algorithms over 100 random samples of BSs-user
placements.

\medskip\noindent The different simulations on \Cref{fig:plots50,fig:plots200}
show, that our heuristic \textsc{DP-Similarity Clustering} algorithm provides
high quality solutions with low time complexity for the total interference
minimization problem in bipartite graphs. Applying it for typical wireless
networks, it nicely optimizes the total interference in the overall
communication network. Compared with the state-of-the-art spectral clustering
method, it is clear that the proposed algorithm achieves better performance with
much less (computational) complexity.

For the somewhat unrealistic choice of $\alpha=1$, the performance of
\textsc{Spectral Clustering} is more stable, and in certain scenarios, its
performance even surpasses that of
\textsc{DP-Similarity Clustering}: for example, for $b=200$ BSs $u=500$ users, and
$M\le 25$.

\subsection{Analysis of the running times}\label{sec:runningtimes}
If the order of magnitude of $M$ is reasonable, i.e., $b-M=\Omega(b)$, then
the running times of neither \textsc{DP-Similarity Clustering} nor \textsc{Spectral
Clustering} depend too much on the exact value of $M$; in fact, the difference
in running time between $M=20$ clusters and $M=40$ clusters with $b=200$ BSs and
$20\le u\le 500$ users is about $5\%$. Thus the displayed running times are
shown for reasonable values of $M$.

Theoretically, the running time of \textsc{DP-Similarity Clustering} is
dominated by the matrix multiplication ${W\cdot W^T}$ (computing the similarity
measure), but this operation takes less than
$0.5\,\mathrm{ms}$ even for a $200\times 500$ matrix (thanks to the accelerated
vector operations in the x86-64 instruction set). In this regime of $b\le 200$,
most of the running time of \textsc{DPH-clustering} (\Cref{alg:clusteringB}) is
spent after the matrix multiplication, which is log-quadratic in the number of
base-stations, but the constant factor of the main term is probably quite large.
Thus, to improve the efficiency of the \textsc{DP-Similarity Clustering} algorithm,
further development should be focused on \textsc{DPH-clustering}.

The running time of \textsc{Spectral Clustering} is a combination of computing the
eigenvectors of a matrix of order $b+u$ and subsequently clustering the
normalized eigenvectors via the $k$-means algorithm. Considering the large
base matrix and the complex operations performed on it, it is not surprising
that \textsc{Spectral Clustering} is an order of magnitude slower than
\textsc{DP-Similarity Clustering}.

Let us refer back to our considerations after \Cref{th:interference}.
In real-world applications, users are constantly entering and leaving network,
and some of them may move out of the range of their cells. For this reason, the
clustering needs to be frequently updated, but the time complexity may be
prohibitive if the number of BSs is very large. However, if the overall changes
to $W$ are not large, it is possible to reuse $\mathcal{I}_{|I|-M}$, and thus
\textsc{DPH-clustering} need not be called every time $G$ or $W$ is updated.
Practically, \textsc{DPH-clustering} is called only if the $\mathrm{tinf}$ value
of the solution increases beyond a preset threshold. Since users are joined to
the respective best cluster by \Cref{alg:dpBbestU}, if the users know the current
$\mathcal{I}_{|I|-M}$, they can decide themselves individually (locally)
when to leave and join another cell.

\section{Conclusions}
We have proposed a robust and deterministic algorithm to solve the total
interference minimization problem. We have demonstrated through analysis and simulation
that it provides higher quality solutions than the popular
\textsc{Spectral Clustering} method. The algorithm runs quickly
enough to be considered in real-world applications.

Next, we list two suggestion for future research problems, whose solutions can
considerably increase the usefulness of our algorithm in the wireless network
domain.
	
The first problem is this: suppose that the clustering problem is restricted to
such that any cluster may contain at most $T$ BSs; assume also that we want to
have (at most) $M$ clusters in the solution.
However, prescribing both the upper bounds $M$ and $T$ may prevent the
existence of feasible solutions, for example if the number of base-stations is
more than $M \times T$. It is easy to construct
examples where the greedy \textsc{DPH-clustering} algorithm
(\Cref{alg:dpBbestU}) will not be able to satisfy the two conditions
simultaneously, even though many feasible solutions exist. It is very probable
that some form of backtracking capability could help tremendously,
but we have not tried to address this problem yet.

A more particular problem can be described as follows: our proposed algorithm assigns
every BS to some cluster in the total interference minimization problem. However
it is possible that using a certain BS in any cluster causes more interference
than not using it at all. This problem can be dealt with a trivial
post-processing procedure: after the clustering $\mathcal{P}=\{P_1,\ldots,P_M\}$
is determined, delete a tower $i$ from $P_k$ if the removal decreases
$\bar{w}(P_k)/{w(P_k)}$, since removing a BS from a cluster cannot increase the
interference fractions of other clusters.

\section*{Declarations}
\noindent \textbf{Conflict of interest}: The authors declare that they have no conflicts of interest to report regarding the present study

\bibliographystyle{informs2014}
\bibliography{references.bib}

\ACKNOWLEDGMENT{%
The authors were supported in part by the National Research,
    Development and Innovation Office -- NKFIH grant SNN~135643, K~132696.\\
We would like to thank Rolland Vida (senior member,~IEEE) for his helpful remarks regarding the writing and organization of this paper.
}%

\end{document}